\documentclass{article}
\usepackage{epsf}
\usepackage{dcolumn}
\usepackage{array}

\newcommand{\pla}[3]{    {\it Phys. Lett. }{ A \bf{#1},} (#3) #2 }

\newcommand{\mpla}[3]{    {\it Mod. Phys. Lett. }{A \bf{#1},} (19#3) #2 }
\newcommand{\eq}[1]{{eq.~(\ref{#1})}}

\newcommand{\etal}{{\it et al.}}

\newcommand{\bea}{\begin{eqnarray}}
\newcommand{\beq}{\begin{equation}}
\newcommand{\eea}{\end{eqnarray}}
\newcommand{\eeq}{\end{equation}}
\newcommand{\nnu}{\nonumber}
\begin{document}
%%%%%%%%%%%%%%%%%%%%%%%%%
\title{Symmetric Triple Well with Non-Equivalent Vacua:
Simple Quantum Mechanical Approach }
{\small
\author{H. A. Alhendi$^1$\ and \ E. I. Lashin$^{1,2}$\\
$^1$ Department of physics and Astronomy, College of Science,\\ King Saud University, Riyadh,
Saudi Arabia \\
$^2$ Department of Physics, Faculty of Science, \\Ain Shams University, Cairo, Egypt}
}
\maketitle
\begin{abstract}
The structure of the energy levels in a deep triple well is analyzed
using simple quantum mechanical considerations. The resultant
spectra of the first three energy levels are found to be composed of a ground
state localized at the central well and the two other states are
distributed only among the left and right well in anti-symmetric and symmetric
way respectively. Due to the tunneling effects the energy eigenvalue of
the ground state is approximately equal to the ground state energy for a harmonic
oscillator localized at the central well, while the two others
are nearly degenerate with approximate values equal to the
ground state energy of a harmonic oscillator localized at the left
or right well. The resulting pattern of the spectra are
confirmed numerically. The failure of the instantonic
approach recently applied for predicting the correct spectra is commented. 
\end{abstract}

Understanding the structure of the energy levels in a triple well is
important from both theoretical and practical point of view and quite different from
the symmetric double well.
In this work we consider the triple well of the form;
\beq
V(x)={\omega^2 \over 2}\, x^2\left(x^2-1\right)^2,
\label{pot}
\eeq
The Schrodinger equation corresponding to the potential in
\eq{pot} ($\hbar=1 , m=1$) reads
\beq
-{1\over2} {d^2\Psi\over d x^2} +{\omega^2 \over 2}\,
x^2\left(x^2-1\right)^2\Psi = E \Psi
\label{sch}
\eeq
%%%%%%%%%%%%%%%%%%%%%%%%%%%%%%FIGURE.1%%%%%%%%%%%%%%%%%%%%%%%%%%%%%%%%%%%
%%%%%%%%%%%%%%%%%%%%%%%%%%%%%%%%%%%%%%%%%%%%%%%%%%%%%%%%%%%%%%%%%%%%%%%%%%%
\begin{figure}[htbp]
\epsfxsize=7cm
\centerline{\epsfbox{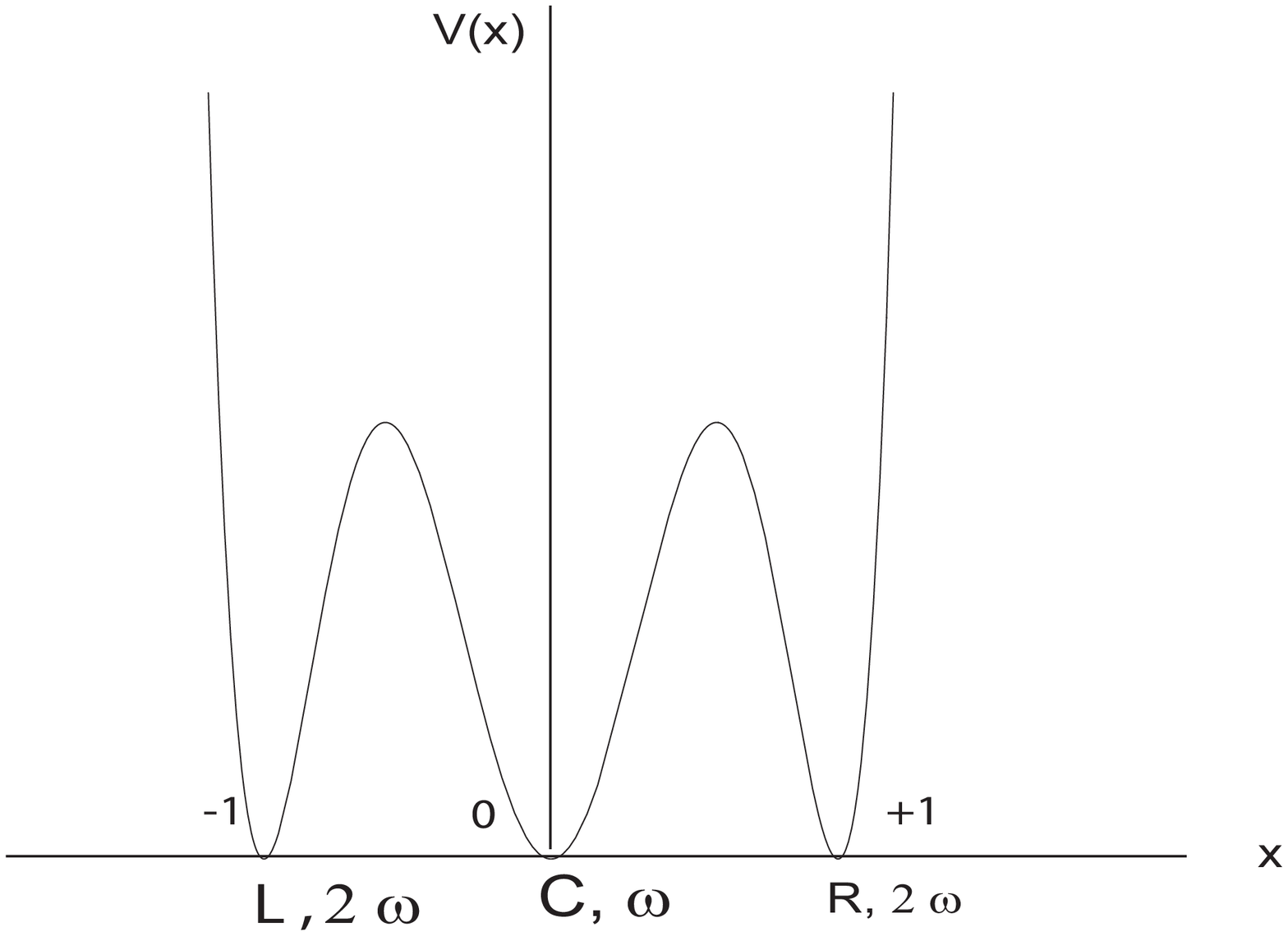}}
\caption{\footnotesize
Triple well: $V(x)={\omega^2 \over 2}\, x^2\left(x^2-1\right)^2$}
\label{triplewell}
\end{figure}
%%%%%%%%%%%%%%%%%%%%%%%%%%%%%%%%%%%%%%%%%%%%%%%%%%%%%%%%%%%%%%%%%%%%%
%%%%%%%%%%%%%%%%%%%%%%%%%%%%%%%%%%%%%%%%%%%%%%%%%%%%%%%%%%%%%%%%%%%%
The potential ,$V(x)$, shown in fig.~\ref{triplewell}, consists of
central, left and right well separated by barriers. In the limit
of highly enough barriers (deeply enough wells), the low lying
energy states can be described as linear combination of harmonic
oscillator states localized over the wells at $x=0,\pm 1$. In the
vicinity of each well the potential can be approximated as a
harmonic oscillator potential. The frequency corresponding to the
central well is $\omega$, while for the left and right well is $2\,\omega$
as can be easily verified by expanding the potential around each
minima. To fix our notations the state $|C\rangle$ denotes the harmonic oscillator ground
state  confined to the central well, while $|L\rangle$
and $|R\rangle$ denote the ground states for those confined to
right and left well respectively.

The low lying spectra (first three levels) of the triple well can be
analyzed from simple quantum mechanical considerations. First;
each eigen-state must have a definite parity since the potential in \eq{pot}
admits symmetry under reflection $(x\rightarrow -x)$. Second;
having a trial wave function for a quantum mechanical system
containing some parameters, these parameters can be fixed by using
the variational method.

The most general linear combination of harmonic oscillator states
forming the ground and the first two excited states are the
following,
\bea
|\Psi_0\rangle &=& \alpha_0 |R\rangle + \beta_0 |C\rangle +
\gamma_0 |L\rangle , \nnu\\
|\Psi_1\rangle &=& \alpha_1 |R\rangle + \beta_1 |C\rangle +
\gamma_1 |L\rangle , \nnu\\
|\Psi_2\rangle &=& \alpha_2 |R\rangle + \beta_2 |C\rangle +
\gamma_2 |L\rangle , \nnu\\
\label{lincom}
\eea
where $\alpha's$, $\beta's$ and $\gamma's$ are free parameters are
taken to be real.

Parity requirement namely,
\bea
{\cal P}|\Psi_0\rangle = |\Psi_0\rangle ,
&{\cal P}|\Psi_1\rangle = -|\Psi_1\rangle ,
&{\cal P}|\Psi_2\rangle = |\Psi_2\rangle ,
\label{par1}
\eea
where $\cal{P}$ is the parity operator, allows to take the parity for
the states $|C\rangle, |R\rangle , \mbox{and} |L\rangle$ as
\bea
{\cal P}|R\rangle = |L\rangle ,&{\cal P}|C\rangle = |C\rangle ,
& {\cal P}|L\rangle = |R\rangle .
\label{par2}
\eea
The other requirements are orthogonality and normalization relations of the
states namely
\beq
\langle\Psi_i|\Psi_j\rangle = \delta_{i,j}\hspace{1cm},\hspace{1cm} i,j=0,1,2
\label{ortho}
\eeq

Taking into account the previously mentioned requirements  we get
\beq
\alpha_0=\gamma_0,\;\; \alpha_1=-\gamma_1,\;\; \alpha_2 = \gamma_2,
\;\; \beta_1=0, \;\; \alpha_1={1\over \sqrt{2}},
\label{spec1}
\eeq
and the following equations
\bea
2\alpha_0^2 + \beta_0^2 &=& 1,\nnu\\
2\alpha_2^2 + \beta_2^2 &=& 1,\nnu\\
2\alpha_0\,\alpha_2 + \beta_0 \,\beta_2 &=& 0 .\nnu\\
\label{spec2}
\eea
Introducing angle parameterizations
\bea
\sqrt{2}\,\alpha_0 = \sin{\theta} & ,& \beta_0 = \cos{\theta},
\nnu\\
\sqrt{2}\,\alpha_2 = \cos{\phi} &, & \beta_2 = \sin{\phi},
\nnu\\
\label{angle1}
\eea
we find the first two relations of \eq{spec2} are satisfied
identically and the last relation implies
\beq
\sin{(\theta + \phi)}=0 \Rightarrow \theta + \phi = \pm n
\pi ,\;\;\; n=0,1,2,\cdots ,
\label{angle2}
\eeq
taking the solution $\theta + \phi = \pi$,
the resultant states are
\bea
|\Psi_0\rangle &=& {1\over \sqrt{2}}\left(\sin{\theta}\,|R\rangle +
\sqrt{2} \cos{\theta}\, |C\rangle +
\sin{\theta}\, |L\rangle \right) \nnu\\
|\Psi_1\rangle &=& {1\over \sqrt{2}}\left(|R\rangle -|L\rangle \right)\nnu\\
|\Psi_2\rangle &=& {1\over \sqrt{2}}\left(-\cos{\theta}\,|R\rangle
+ \sqrt{2} \sin{\theta}\,|C\rangle -
\cos{\theta}\,|L\rangle \right) \nnu\\
\label{lincom2}
\eea

The energy for each state given by eqs.~\ref{lincom2},
obtained from the expectation value of the corresponding
Hamiltonian, are
\bea
\langle\Psi_0| H |\Psi_0\rangle&=& {1\over 2}\left(\omega_1 \sin^2{\theta} + \omega_0
\cos^2{\theta}\right)\nnu \\
\langle\Psi_1| H |\Psi_1\rangle &=& {\omega_1 \over 2} \nnu \\
\langle\Psi_2| H |\Psi_2\rangle &=& {1\over 2}\left(\omega_1 \cos^2{\theta} + \omega_0
\sin^2{\theta}\right),\nnu \\
\label{ener1}
\eea
where $\omega_0$ (corresponding to $\omega$) and $\omega_1$
(corresponding to $2\,\omega$)
are the angular frequencies of the harmonic
oscillator near the vicinities of central and left(right) well
respectively. Minimizing the energy of the ground state with
respect to the angle parameter $\theta$, we find that the minimum
occurs at $\theta = n\,\pi$ for $\omega_1 > \omega_0$. The the energy eigenvalues are
then:
\bea
E_0&=&{\omega_0 \over 2}={\omega \over 2},\nnu\\
E_1 = E_2 &=& {\omega_1 \over 2}=\omega ,\nnu \\
\label{ener2}
\eea
The results here are for the ideal case -The limits of very deeply enough well without
tunneling- but due tunneling effect,
the first two excited states become nearly degenerate with energy
approximately equal to $\omega$ while the ground state energy
has approximately the value  ${\omega \over 2}$.

These energy pattern spectra are inconsistent with the results
based on instantonic approach in Refs.~\cite{casa,lee}, which
predicted that the first three levels are equidistant and nearly
degenerate with midpoint energy equals to ${3\, \omega \over 2}$.
It seems to us that this resulted from an improper application of
the instanton method.
In contrast the simple quantum mechanical
consideration presented in this work implies that  the first
and second excited states are nearly degenerate with
a finite non vanishing energy gab separating them from the ground
state as given by \eq{ener2}.

\begin{figure}[hbtp]
\centering
\begin{minipage}[c]{0.33\textwidth}
\epsfxsize=4cm
\centerline{\epsfbox{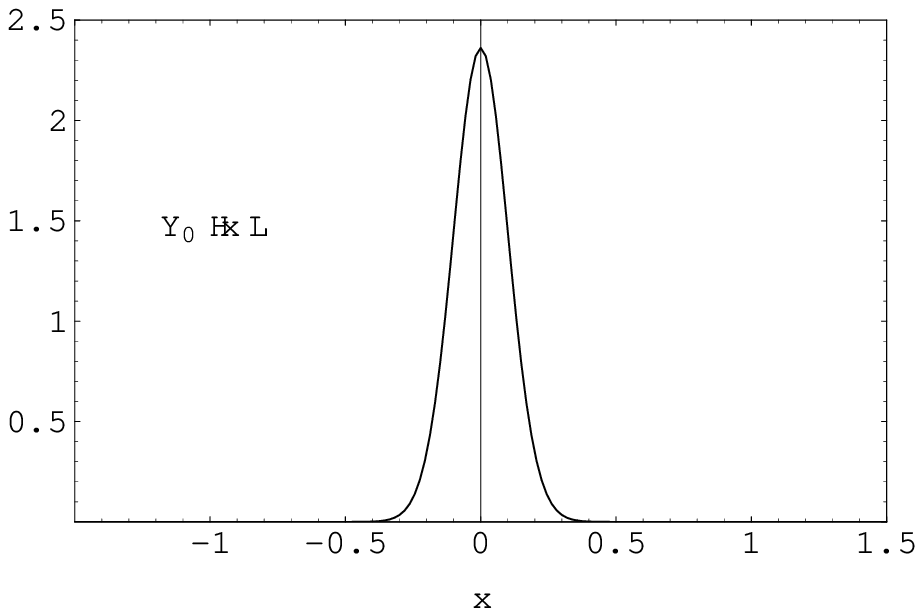}}
\end{minipage}%
\begin{minipage}[c]{0.33\textwidth}
\epsfxsize=4cm
\centerline{\epsfbox{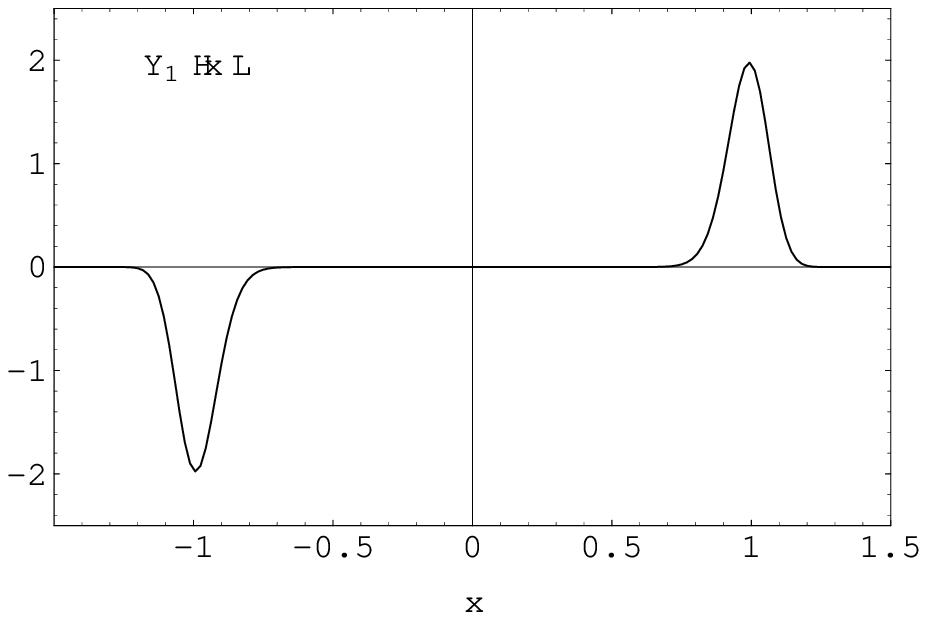}}
\end{minipage}%
\begin{minipage}[c]{0.33\textwidth}
\epsfxsize=4cm
\centerline{\epsfbox{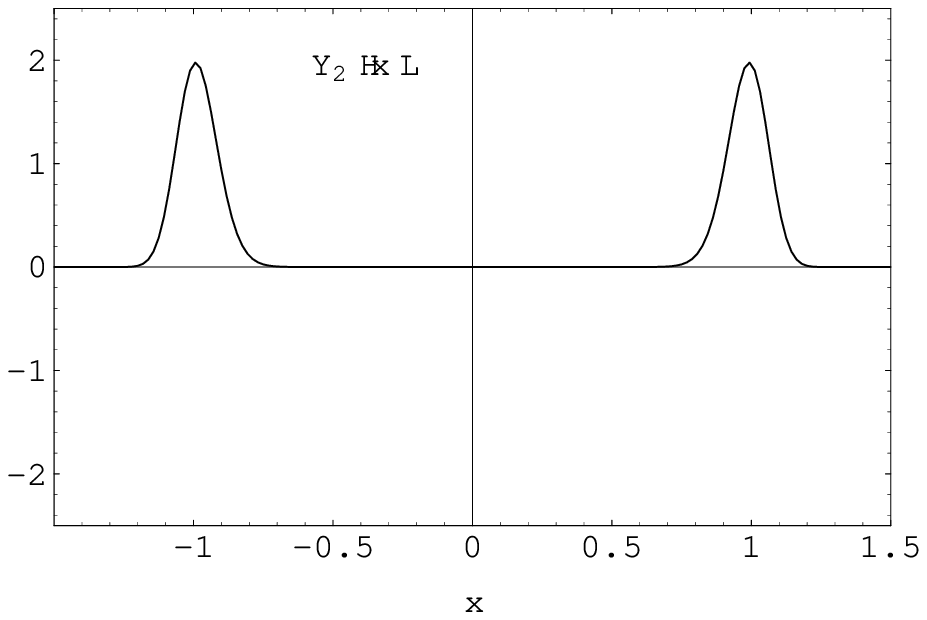}}
\end{minipage}
\caption{\footnotesize The normalized ground (left), first excited (middle)
and second excited (right) state wave functions for $\omega=100\ \mbox{and}, L={3\over 2}$
for the bounded triple well.}
\label{figtriple}
\end{figure}

To justify the results derived  here from simple quantum mechanical
considerations, we resort to numerical solution for Schrodinger
equation.
For this purpose we apply the power series method in a finite range, of width $2\,L$,
for \eq{sch}.
This method has been previously
applied to variant potential functions with and without
degenerate minima leading to results with high accuracy
\cite{ourdw,ourmw}, and its details can be found in Ref.\cite{ourdw}.
In this short letter we present only
\begin{table}[hbtp]
\begin{tabular}{l|ccc}
\hline
\hline
$\omega , L , I$ & $20, 2, 750$ & $40, 2, 1000$ & $60, 1.5 , 1000$ \\
\hline
\hline
$E_0$ & 9.1100715702553 &19.200084475112926 &29.218766418207469 \\
$E_1$ &17.5140977513941 &37.948103273585804 &58.017242546933332 \\
$E_2$ &17.6975924458074 &37.948176236685948 &58.017242556963103 \\
\end{tabular}
\begin{tabular}{l|ccc}
\hline
\hline
$\omega , L , I$ & $80, 1.5 , 1000$ & $100,1.5 , 1500$ &  \\
\hline
\hline
$E_0$ &39.227231934365212 &49.23207941514294132072024439682 &\\
$E_1$ &78.047249798583686 &98.06414003277967330221270153882 &\\
$E_2$ &78.047249798584662 &98.06414003277967338155670061896 &\\
\end{tabular}
\caption{\footnotesize The numerically calculated energy eigenvalues for the first three levels. The
numbers $\omega , L , I$ represent the parameter appearing in the triple well
potential, the width between the two infinite walls, and the number of non vanishing
terms in truncated power series solution respectively}
\label{tab1}
\end{table}
the numerical results for the energy eigenvalues and eigen-states in
Table~\ref{tab1} and Fig.~\ref{figtriple} respectively.
these reveal that low lying spectra (first three
levels) consist of combinations of harmonic oscillator states
localized at the central, left and right well. More precisely, the
ground state is the ground harmonic oscillator state localized at
the central well. While the first and second excited states are
linear combination of the ground harmonic oscillator states
localized at the left and right well having odd and even parity
respectively. These results assert the implication presented in this
work, and contradict that obtained in Refs.\cite{casa,lee}. The
method here, can be extended to higher excited states and
multi-well potentials.
We hope the present work will encourage for a proper treatment of the
instantonic methods for potential function with non equivalent
vacua.
%%%%%%%%%%%%%%%%%%%%%%%%%%%%%%%%%%%%%%%%%%%
%%%%%%%%%%%FIG. 2 %%%%%%%%%%%%%%%%%%%%%%%
%%%%%%%%%%%%%%%%%%%%%%%%%%%%%%%%%%%%%%%%%%%%%%
%%%%%%%%%%%%%%%%%%%%%%%%%%%%%%%%%%%%%%%%%%%%%%%
\section*{Acknowledgement}
This work was supported by research center at college of science,
King Saud University under project number Phys$/1423/02$.


\begin{thebibliography}{99}
{\small
\bibitem{casa} J. Casahorr$\acute{a}$n, \pla{283}{285}{2001}.
\bibitem{lee} S. Y. Lee \etal,
\mpla{12}{1803}{97}.
\bibitem{ourdw} H. A. Alhendi and E. I. Lashin {\it
quant-ph/0305128, submitted to Cand. J. Phys.}
\bibitem{ourmw} H. A. Alhendi and E. I. Lashin {\it
quant-ph/0306016}.
%%%%%%%%%%%%%%%%%%%%%%%%%%%%%%%%%%%%%%%%%%%%%%%%%%%%%%%%%%%%%%%%%%%%%%%%
}
\end{thebibliography}
\end{document}